\documentclass[review]{elsarticle}
\usepackage{lineno,hyperref}
\modulolinenumbers[5]

\biboptions{authoryear}

\def\({\left(}
\def\){\right)}
\def\[{\left[}
\def\]{\right]}

\usepackage{color}

\begin{document}

\begin{frontmatter}

\title{Accretion around black holes: The geometry and spectra}


\author[address1,address2]{B.F. Liu\corref{mycorrespondingauthor}}
\ead{bfliu@nao.cas.cn}
\author[address1,address2]{Erlin Qiao}
\ead{qiaoel@nao.cas.cn}
\cortext[mycorrespondingauthor]{Corresponding author}

\address[address1]{Key Laboratory of Space Astronomy and Technology, National Astronomical 
Observatories, Chinese Academy of Sciences, Beijing 100012, China}
\address[address2]{School of Astronomy and Space Sciences, University of Chinese Academy of 
Sciences, 19A Yuquan Road, Beijing 100049, China}

\begin{abstract}
Observations of black hole X-ray binaries and active galactic nuclei indicate that the accretion 
flows around black holes are composed of hot and cold gas, which have been theoretically described 
in terms of either a hot geometrically thick corona lying above and below a cold geometrically 
thin disk or an inner advection dominated accretion flow connected to an outer thin disk. 
This article reviews the accretion flows around black holes, with an emphasis on the physics that 
determines the configuration of hot and cold accreting gas, and how the configuration varies 
with the accretion rate and thereby produces various luminosity and spectra.  
\end{abstract}


\end{frontmatter}


\section{Introduction: solutions of accretion flow}

It has been widely accepted that gas accretion onto black holes provides the principal power 
of radiations from black hole X-ray binaries (BHXRBs) and active galactic nuclei (AGNs). 
With the action of viscosity, gas captured by the black hole spirals toward the central 
black hole, drops off its initial angular momentum outwards, releases gravitational potential 
energy into heat. A fraction or all of the viscous heat is radiated, yielding a spectrum in 
dependence on  specific radiation processes. The accretion flows are described by four 
basic solutions,  that is,  the standard thin disk \cite[e.g.][]{Shakura1973A&A....24..337S}, 
the optically thin two-temperature disk  \cite[][]{Shapiro1976ApJ...204..187S}, the slim 
disk \cite[e.g.][]{Katz1977ApJ...215..265K,Begelman1978MNRAS.184...53B,Abramowicz1988ApJ...332..646A} 
and the advection-dominated accretion 
flow \cite[e.g.][]{Ichimaru1977ApJ...214..840I,Rees1982Natur.295...17R,Narayan1994ApJ...428L..13N,
Narayan1995ApJ...444..231N,Narayan1995ApJ...452..710N,Abramowicz1995ApJ...438L..37A,
Abramowicz1996ApJ...471..762A}. 
When these solutions are applied to black holes, a composite of them   are commonly adopted. 

The most famous solution is the standard thin disk model developed in 1970s 
by \cite{Shakura1973A&A....24..337S}, 
\cite{Novikov1973blho.conf..343N}, 
\cite{Lynden-Bell1974MNRAS.168..603L}\cite[for textbooks see][]{Frank2002apa..book.....F, Kato2008bhad.book.....K}. The thin disk is a  geometrically thin, optically thick accretion flow, 
characterized by multi-color blackbody radiation with effective temperatures ranging 
from $10^5$K to  $10^7$K from supermassive black holes to stellar-mass black holes.  
The thin disk model has been the common paradigm in accretion theory and  successfully used 
to model a large number of astrophysical systems. The most successful applications are  
the steady state and time-dependent nature of thin disk in dwarf novae and soft X-ray transients. 

The second solution was presented by \cite{Shapiro1976ApJ...204..187S}, 
in which the accreting gas  is optically thin, hot plasma with temperature near virial in ions 
and $\sim 10^9$K in electrons. Such a solution is, however, thermal 
unstable \cite[][]{Piran1978ApJ...221..652P}, and is therefore not considered as applicable to 
black holes.

In the above two solutions, the energy released by viscous heating is assumed to be emitted locally. 
In contrast, a third solution was proposed  \cite[][] {Ichimaru1977ApJ...214..840I,Rees1982Natur.295...17R,Narayan1994ApJ...428L..13N} and extensively investigated \cite[e.g.][]{Narayan1995ApJ...444..231N,Narayan1995ApJ...452..710N,Abramowicz1995ApJ...438L..37A} \cite[see the review][]{Yuan2014ARA&A..52..529Y} by taking into account the advection of viscous heating.  In this solution, the accreting gas has a very low density and the Coulomb coupling between the ions and electrons is weak. The viscous energy can not be radiated efficiently, instead, part of it is stored in the gas as thermal energy and  eventually advected onto the black hole. Therefore,  the gas forms an optically thin, two-temperature accretion flow with the ion temperature close to virial  temperature and electron temperature $T_{\rm e} \sim 10^9$K, referred to as advection dominated accretion flow (ADAF). Apparently, the ADAF solution is only valid at low accretion rates so that the advection plays an important role in "digesting" the viscous heating. The radiation is, therefore, inefficient,  however,  contributing to hard X-ray by synchrotron self-Compton process. 

The fourth solution is developed for the high accretion rate, in which the gas density is very large  \cite[e.g.][]{Katz1977ApJ...215..265K,Begelman1978MNRAS.184...53B,Abramowicz1988ApJ...332..646A,Honma1996PASJ...48...77H,Chen1993ApJ...412..254C}. While  the viscous heating is efficiently transferred to emissions locally, most of photons are trapped in the disk,  not able to escape from the disk surface during the accretion time.   The energy  is largely advected into the black hole in the form of photons rather than thermal energy in the ADAF solution. Therefore, the energy conversion into radiation  is inefficient too.   The temperature in this solution  is higher than that in a standard thin disk as a result of larger accretion rate, while it is lower than that in an ADAF since ions can easily transfer viscous heat to electrons through Coulomb collisions. The  disk is radiation-pressure supported, thereby the vertical scale height is larger than a standard disk, but not so extensive as an ADAF.  The solution is, thus, called slim disk, sometimes  referred to as an optically thick, advection-dominated accretion flow.

Other solutions of the accretion flows, the adiabatic inflow-outflow solution (ADIOS), 
convection-dominated accretion flow (CDAF), and the luminous hot accretion 
flow (LHAF) are the variants of ADAF,  emphasizing the roles of  outflows,  
convection and radiation efficiency, respectively. The ADIOS, CADF, and LHAF are not 
discussed as particular solutions in addition to the ADAF.
 
Application of the four accretion models to black holes is constrained by both the 
theoretical assumption of specific models and observational luminosity and spectrum.  
The thin disk model applies for objects with  luminosity below the Eddington luminosity,  
$ L_{\rm Edd}= {4\pi GMm_p c\over \sigma_T}$,  the critical luminosity for the 
outward radiation force (through Thomson scatterings) to balance the central gravity. 
For objects with high luminosity, slim disk model is often adopted. Examples are the  
ultra-luminous X-ray sources (ULXs), the narrow-line Seyfert 1 galaxies (NLS1s), 
and the tidal disruption events. On the other hand, the ADAF model, 
characterized by its inefficient  radiation and high temperature, successfully applies 
to the quiescent and low state BHXRBs, the Galactic center black hole, Sgr A$^*$, and 
the low-luminosity AGNs.
   
The luminosity $L$ is translated to the accretion rate $\dot M$ with 
$\dot M=L/ \eta c^2$, where $\eta$ is 
the  radiation efficiency.  Accordingly, the Eddington ratio $L/ L_{\rm Edd}$  is  related  to 
the Eddington scaled accretion rate $\dot m$, $L/ L_{\rm Edd}=(\eta/ 0.1)\dot m$, 
with $\dot m\equiv\dot M/ \dot M_{\rm Edd}$ and $\dot M_{\rm Edd}\equiv L_{\rm Edd}/ 0.1 c^2$.
By this definition, the thin disk model applies for  $\dot m \le 0.3$.
For higher accretion rates,  $\dot m > 0.3$, the accretion is via a slim disk, while for lower accretion rates, 
$\dot m<(0.1-0.3) \alpha^2 $ \cite[e.g.][]{Narayan1995ApJ...452..710N,Yuan2014ARA&A..52..529Y}, 
the accretion is via an ADAF according to the "strong ADAF principle" \cite[][] {Narayan1995ApJ...452..710N}, 
though the thin disk is valid in this regime.

\section{Theory confronted with observation: configuration of accretion flows around black holes}
Black hole X-ray transient binaries  can be used as a probe of the accretion process over a wide range in luminosity. The X-ray spectral behavior of these systems exhibits different states and transitions, in particular, the X-ray spectral state transition between the well-known high/soft state and low/hard state \cite[see the reviews ][] {Tanaka1996ARA&A..34..607T, Remillard2006ARA&A..44...49R}.  Combined with the optical/UV observations, the X-ray observations reveal that  these two spectral states originate from different accretion modes. At low luminosities the systems are characterized by  a power law spectrum in hard X-rays, accompanied with a weak thermal optical component. The X-ray emission is commonly thought to be produced by the Compton scattering of soft photons off the hot electrons in an inner ADAF and the optical component by  an outer truncated thin disk \cite[e.g.][]{Lasota1996A&A...314..813L, Dubus2001A&A...373..251D,  Poutanen2018A&A...614A..79P}. 
At high luminosities the systems are characterized by a multicolor blackbody component that dominates at about 1 keV, accompanied with a weak hard tail sometimes. This has been interpreted as arising from a geometrically thin, optically thick accretion disk, covered by a weak, hot corona \cite[e.g.][]{Poutanen1998PhST...77...57P,Gierlinski1999MNRAS.309..496G}. The transition occurs at  an Eddington ratio around $\sim 0.02$ \cite[e.g.][]{ Maccarone2003A&A...409..697M, Gierlinski2006MNRAS.370..837G}, though there is large difference between individual sources. 
 
Broadband spectral features of AGNs also provide strong evidence for hot gas coexist with the thin disk  in the neighborhood of the supermassive black holes \cite[e.g.][]{Mushotzky1993ARA&A..31..717M, Zdziarski1999MNRAS.303L..11Z}.  
The big blue bump and the power-law X-ray radiation are  commonly thought to be the disk thermal emission and inverse Compton scattering of the disk photons  with the high-temperature electrons  in the hot corona, respectively  \cite[e.g.][]{ yuan2010}. The fluorescent iron lines are explained by reflection of the  corona radiation by the underlying disk \cite[e.g.][]{Fabian2000PASP..112.1145F, Reeves2001A&A...365L.134R}, supporting the cold disk and hot corona scenario.

 Observational data of black holes in recent years display more complicated spectral features, indicating a combination of various  accretion flows \cite[e.g.][]{You2021NatCo..12.1025Y,Kara2019Natur.565..198K,Ruan2019ApJ...883...76R,Jin2017MNRAS.468.3663J,Jin2017MNRAS.471..706J},   accompanied by wind/outflows in some cases \cite[for observational and theoretical studies of the wind see e.g. ][]{Tombesi2010,Gofford2015,Zhu2018,Nomura2016,Yang2018,Yang2019,Wang2013Sci,Shi2021,Yuan2015,Bu2016,Bu2018}.    
A common feature of the accretion geometry in all these observations is the co-existence of hot and cool accretion flows, of which the well-known  composites  are either an inner ADAF connecting to a truncated disk, or an accreting corona lying above a standard thin disk which extends inward to the innermost stable circular orbit  (ISCO). The former configuration was proposed by \cite{Esin1997ApJ...489..865E} and applied to both quiescent BHXRBs and low-luminosity AGNs, and the latter configuration was proposed to explain the power-law X-ray emission and lower frequency blackbody component observed in both BHXRBs and AGNs  \cite[][] {Bisnovatyi-Kogan1976SvAL....2..191B, Liang1977ApJ...218..247L,Haardt1991, Haardt1993,Nakamura1993, Svensson1994, Zdziarski1999MNRAS.303L..11Z}.   
 
The  physical mechanism triggering the transition between such configurations has been studied since different accretion solutions were found.  The “strong ADAF principle” suggests that, whenever the accreting gas has a choice between a standard thin disk and an ADAF, the ADAF configuration is chosen  \cite[][] {Narayan1995ApJ...452..710N}. With the radius-dependent critical accretion rate for existence of the ADAF, this rule provides a transition radius  for a given accretion rate \cite[e.g.][]{Esin1997ApJ...489..865E}. In a different approach \cite{Honma1996PASJ...48...77H} determines the truncation radius by considering radial diffusive transport of heat between the inner hot  accretion flows and the outer cool disk . The thermal instability in the radiation pressure-supported disk can also trigger a transition from a thin disk to an ADAF in the inner region  \cite[][] {Gu2000ApJ...540L..33G, Lu2004ApJ...602L..37L}.  Among various possibilities,  the disk and corona interaction model provides a promising explanation for the disk truncation and the spectral state transition  \cite[][] {liu1999, Meyer2000a, Meyer2000b,  Rozanska2000aA&A...360.1170R, Rozanska2000bMNRAS.316..473R, Meyer2001a, Meyer2002, liu2002a, Spruit2002A&A...387..918S, Meyer2003, liu2003, Dullemond2005A&A...434..415D, liu2005,liu2006a, liu2007, Meyer2007, Taam2008, Meyer2009, liu2009, liu2011,liujy2012, qiao2012,Taam2012,qiao2013a,qiao2013b,liu2013a, liu2013b,liu2015,liujy2016, liu2017,Meyer2017, qiao2017,qiao2018a, Taam2018, qiao2020b}. Specifically, the interaction between the disk and corona causes disk gas evaporating to the corona or coronal gas condensing to the disk, depending on the gas supply rate and how the gas feeds to the accretion.  In the case that the  accreting gas is supplied to the thin disk, as in low-mass X-ray binaries of  black hole systems (LMXBs), gas evaporates from the disk to the corona. Thus, the disk is truncated when the gas supply is insufficient for evaporation. However,  it extends to the ISCO when the gas supply is more than the evaporation \cite[e.g.][]{Meyer2000a, Meyer2000b, liu2002a}).  In the case that the accreting gas is hot, as the wind  in high-mass X-ray binaries of black hole systems (HMXBs) or Bondi flows in AGNs, the hot accretion flow  condenses partially to the disk  if the supply rate is high,  while it keeps the pure ADAF form if the gas supply rate is low \cite[e.g.][]{liu2015,qiao2018a,Taam2018}.    

The coexistence of different accretion flows has been found  by magnetohydrodynamic (MHD) simulations \cite[e.g.][]{Miller2000ApJ...534..398M, Hawley2001ApJ...548..348H,McKinney2014MNRAS.441.3177M,Ohsuga2005ApJ...628..368O, Sadowski2016MNRAS.456.3929S}.  In particular, the  two-phase accretion flow, either in a form of a corona overlapping a thin disk or an inner corona connecting outer thin disk,  has been confirmed by the global radiation MHD simulations \cite[e.g.][and references therein]{Jiang2019ApJ...885..144J}. 

 It is now widely accepted that the accretion flows around black holes are composed of hot and cold flows.  The relative strength of the two flows, either  the disk and the corona at high states, or  the inner ADAF and the outer disk at low states is constrained by observational spectral energy distribution (SED). For instance,   in modeling the soft and hard components in BHXRBs, a proper value of  the truncated radius is  adopted \cite[e.g.][and references therein]{Poutanen2018A&A...614A..79P}; While in interpreting the strong X-ray emission in luminous AGN,  a certain fraction of accretion energy is assumed to release in the corona \cite[e.g.][]{Haardt1991,Haardt1993}.   Then, what is the  physical mechanism  that  drives the formation  of  the hot and cold flows?  How does the relative strength change?  Can one determine the truncation radius consistently with observations?  We'll give a detailed review aiming to answer such questions. 
  
 In the following sections,  we describe the interaction between the hot and cold flows, elucidate how the interaction leads to disk truncation and consequently  determines the relative strength of hot and cold accretion flows as a function of black hole mass, accretion rate and viscosity parameter, thereby producing spectra comparable with observations. 
 Specifically, the disk evaporation model applicable to the accretion of Roche-lobe overflow (RLOF) is presented in Sect.\ref{s:evaporation}, and the corona condensation model applicable for accretion of wind is presented in Sect.\ref{s:condensation}.  The geometry of two-phase accretion flows as a consequence of disk-corona interaction is displayed  in Sect.\ref{s:geometry} . The radiative properties of the disk-corona system are shown and compared with observations  in Sect.\ref{s:radiation}. The conclusions are  in Sect.4.

\section{Formation of two-phase accretion flows around black holes}

  The two-phase accretion flows  refer to  the  coexistent cold and hot  accretion flows,  which are, respectively,  the standard thin disk and the ADAF/corona.  The corona is physically similar to the ADAF excepted that a thin disk is sandwiched in. 
 Thus, the properties of the corona, such as temperature and the optical depth, can be strongly 
affected by the cold thin disk  via heat conduction, radiation coupling and the consequent mass exchange at a dynamic equilibrium between the disk and the corona.

The concept of the evaporation of matter from an accretion disk was
originally proposed in the pioneering work by  \cite{Meyer1994} to explain
the  UV lag observed in dwarf novae. The model was developed
in a more detailed form for application to accretion flows surrounding black holes  \cite[][] {liu1999, Meyer2000a, Meyer2000b, Meyer2001a,liu2002a}, which demonstrated the possible truncation of an outer optically thick disk and the
consequent formation of an inner optically thin ADAF. 
The incorporation of various elements was included and the effects  were investigated  \cite[][] {Meyer2002, liu2002a,Meyer2003, liu2005,Meyer2005a,Meyer2006, qian2007,qiao2009}. Similar conceptual models,
but with differences in detail were proposed in a semi-analytical model proposed  \cite[][] {Dullemond2005A&A...434..415D} and in a vertically stratified model \cite[][] {Rozanska2000aA&A...360.1170R, Rozanska2000bMNRAS.316..473R}. 
The coronal condensation, instead of disk evaporation, was found in the innermost region at intermediate accretion rates, leading to a weak inner disk at intermediate state  \cite[][] {liu2006a,liu2007,Meyer2007, Taam2008,Meyer2009, liu2011, Meyer2011, Meyer2012}.  The radiation spectra from such interacting disk and corona  were calculated and  compared with observations in BHXRBs \cite[e.g.][]{qiao2012,qiao2013a,Meyer2014,qiao2015} and AGNs \cite[e.g.][]{liu2009, liujy2012,Taam2012,qiao2013b,liu2013a,liu2013b}.
The condensation model was extended to the case of wind accretion and applied to  HMXBs and AGN  \cite[][] {liu2015, liu2017, Meyer2017, qiao2017, qiao2018a,Taam2018, qiao2019,qiao2020a,Meyer2020}.

\subsection{The disk evaporation model}\label{s:evaporation}

\begin{figure}
\begin{center}
\includegraphics[scale=0.2]{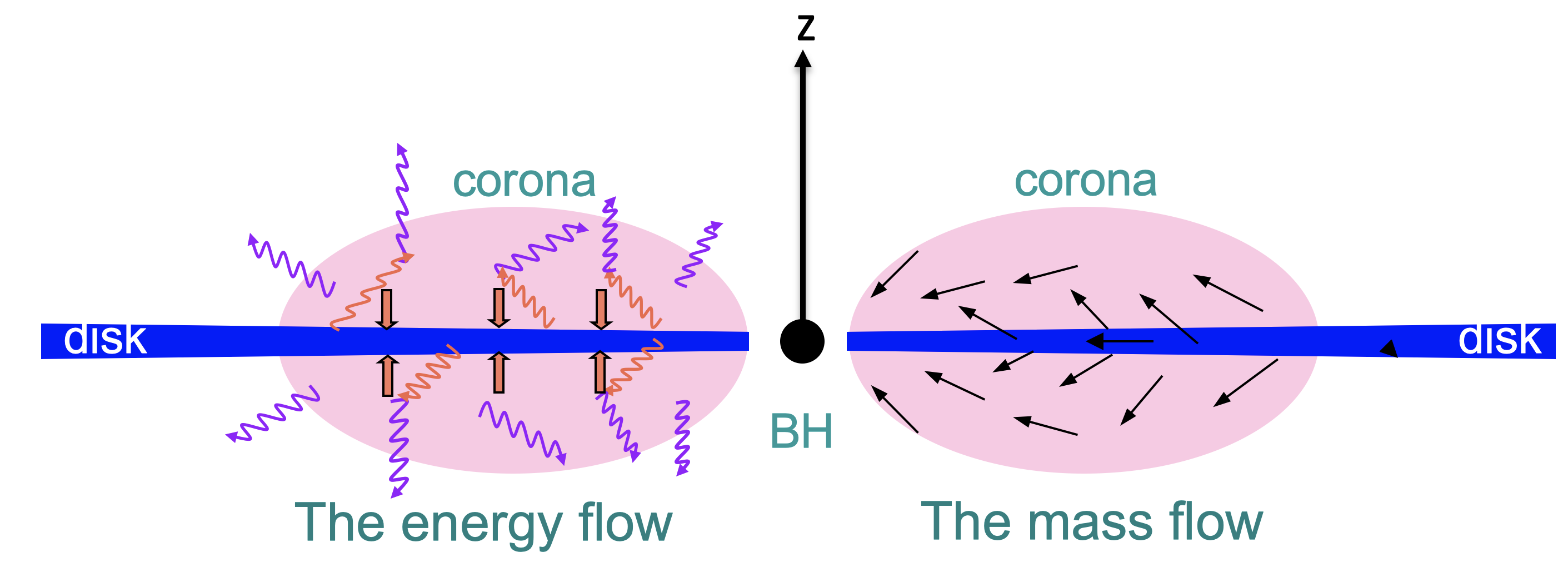}
\caption{Schematic description of the interaction between the disk and the corona. The left part shows the vertical conduction,  the direct and Comption radiations in the corona and the right part shows the gas  flows. }
\label{f:vertical}
\end{center}
\end{figure}

As occurs in LMXBs,  the accretion is assumed to take place via a cold thin disk as gas transferred from the donor
star is constrained to the orbital plane.  A corona  forms above the thin disk either by processes similar to those
operating in the surface of the sun, or by a thermal instability in the uppermost layers
of the disk \cite[e.g.][]{Shaviv1986}. The disk and corona are individually powered
by the release of gravitational energy associated with the accretion of matter, and 
the interaction between the disk and corona is an important and distinctive process.
Specifically, the corona is
an ADAF-like accretion flow modified by the vertical heat conduction and inverse Compton
scattering of disk photons, which play a key role in cooling the electrons as a consequence
of the existence of an underlying disk.  The ions in the corona are directly
heated by viscous dissipation, partially transferring their energy to the electrons by
means of Coulomb collisions. The energy gained by electrons can be effectively 
 conducted to the lower, cooler, and denser  layers by electron-electron
collisions.  In the lower, transition layer, the conductive heat flux is radiated away through
bremsstrahlung only if the number density in this layer reaches a critical value.  If the
density is too low to efficiently radiate the energy, a
certain amount of lower, cooler gas is heated up so that an energy equilibrium is established between the conduction,
radiation and heating of the cool gas. The transfer of gas from the disk to the corona,
to establish the equilibrium, is called evaporation. These processes are schematically described in Figure \ref{f:vertical}.  An opposite process could also take place, that is,  if the density in the transition layer is too high, a certain amount of gas is over-cooled by radiation,  and condenses to the disk (which is described in next subsection).    
The gas evaporating into the corona retains angular momentum and differentially rotates around
the black hole. By frictional stresses, the gas loses angular momentum and drifts inward
in such a way that the corona continuously drains gas towards the black hole. The coronal
accretion flow is re-supplied by continuous evaporation and, therefore, steady flows are
established in the disk and corona with the mass exchange between the two flows.  
The coronal accretion is supplied by evaporating disk matter. Thus, the accretion rate in the corona at a given distance is the sum of the mass evaporation from the outer disk inward to the given distance,  hence it increases toward the black hole as a result of accumulation of the mass evaporation.  When the accretion rate in the corona reaches a critical value,  Coulomb coupling between electrons and ions is efficient so that radiation in the corona is efficient. The increased radiative cooling and decreased conductive heating to the transition layer reduce the evaporation,  thereby evaporation ceases at some radius and condensation takes place in the inner region. Therefore,  there exist a most efficient  evaporation region where the evaporation rate reaches a maximum.   If the gas supply rate to the disk is too low to replenish the disk for the evaporation, the corresponding disk region  will be completely evaporated and filled by hot gas, that is, the disk is truncated.

The corona  temperature, density  and mass flowing rate supplied by the evaporation can be determined by solving the equations describing the dynamics of the corona  \cite[e.g.][]{Meyer2000a,
liu2002a, Meyer2003, qian2007, qiao2009},  as listed below.

Equation of state
\begin{equation}\label{e:EOS}
P={\Re \rho \over 2\mu} (T_{i}+T_{e}).
\end{equation}

Equation of continuity
\begin{equation}\label{e:continuity}
\centering
 {d\over dz}(\rho v_z)= {2\over R}\rho v_R -{2z\over
R^2+z^2}\rho v_z,
\end{equation}
where $2z/(R^2+z^2)$ describes
the gradual expansion of the vertical flow channel with height as the geometry changes
from cylindrical to spherical.

Equation of the $z$-component of momentum
\begin{equation}\label{e:mdot}
\rho v_z {dv_z\over dz}=-{dP\over dz}-\rho {GMz\over
(R^2+z^2)^{3/2}}.
\end{equation}

The energy equation of the ions
\begin{equation}\label{e:energy-i}
\begin{array}{l}
{d\over dz}\left\{\rho_i v_z \left[{v^2\over 2}+{\gamma\over
\gamma-1}{P_i\over \rho_i}-{GM\over (R^2+z^2)^{1\over
2}}\right]\right\}\\
={3\over 2}\alpha P\Omega-q_{ie}\\
+{ }{2\over R}\rho_i v_R
\left[{v^2\over 2}+{\gamma\over \gamma-1}{P_i\over \rho_i}-{GM\over (R^2+z^2)^{1\over 2}}\right]\\
-{2z\over {R^2+z^2}}\left\{\rho_i v_z \left[{v^2\over 2}+{\gamma\over
\gamma-1}{P_i\over \rho_i}-{GM\over (R^2+z^2)^{1\over 2}}\right]\right\},
\end{array}
\end{equation}
where  $q_{ie}$ is the energy exchange rate between the electrons and the ions through Coulomb collisions \cite[][]{Stepney1983,liu2002a},
\begin{equation}\label{e:qie}
{q_{ie}}={\bigg({2\over \pi}\bigg)}^{1\over 2}{3\over 2}{m_e\over
m_p}{\ln\Lambda}{\sigma_T c n_e n_i}(\kappa T_i-\kappa T_e)
{{1+{T_*}^{1\over 2}}\over {{T_*}^{3\over 2}}}
\end{equation}
with
\begin{equation}
T_*={{\kappa T_e}\over{m_e c^2}}\bigg(1+{m_e\over m_p}{T_i\over
T_e}\bigg).
\end{equation}

The energy equation for both the ions and the electrons is
\begin{equation}\label{e:energy-t}
\begin{array}{l}
{\frac{d}{dz}\left\{\rho {v}_z\left[{v^2\over
2}+{\gamma\over\gamma-1}{P\over\rho}
-{GM\over\left(R^2+z^2\right)^{1/2}}\right]
 + F_c \right\}}\\
=\frac{3}{2}\alpha P{\mit\Omega}-n_{e}n_{i}L(T_e)-q_{\rm Comp}\\
+ {2\over R}\rho v_R \left[{v^2\over
2}+{\gamma\over\gamma-1}{P\over\rho}
-{GM\over\left(R^2+z^2\right]^{1/2}}\right]\\
-{2z\over R^2+z^2}\left\{\rho v_z\left[{v^2\over
2}+{\gamma\over\gamma-1}{P\over\rho}-
{GM\over\left(R^2+z^2\right)^{1/2}}\right]
+F_c\right\},
\end{array}
\end{equation}
where $n_{e}n_{i}L(T_e)$ is the bremsstrahlung cooling rate, of which the radiative cooling function
$L(T_e)$  is taken  from  \cite[][] {Raymond1976ApJ...204..290R}; ${q_{\rm {Comp}}}$  is   the  cooling rate through inverse Compton scattering \cite[][]{liu2002a},
\begin{equation}\label{e:comp}
 {q_{\rm Comp}}={4\kappa T_e\over m_e
c^2}n_e\sigma_T c u,
\end{equation}
with $u$ the energy density of the disk radiation; $F_c$ is the thermal conduction flux given by  \cite{Spitzer1962}
\begin{equation}\label{e:fc}
F_c=-\kappa_0T_e^{5/2}{dT_e\over dz}
\end{equation}
with $\kappa_0 = 10^{-6}{\rm erg\,s^{-1}cm^{-1}K^{-7/2}}$ for a fully ionized plasma.

In the above equations, parameters $M$, $P$, $\rho$, $T_i$ and $T_e$ are the mass of black hole, the pressure, density, ion temperature and electron temperature; $n_i$ and $n_e$ are the number density of ions and electrons; $v_z$ and $v_R$  are the vertical and radial speed of the coronal flow;  $\mu=0.62$ is the mean molecular
weight assuming a standard chemical composition (mass fractions of hydrogen and helium are $X=0.75, Y=0.25$) for the corona. 
Constants  $G$, $m_p$ and $m_e$ are respectively is the
gravitational constant, the mass of the proton and the electron; $\kappa$ is the Boltzmann constant, $\Re$ the gas constant, $c$ the speed of light, 
$\sigma$ Stefan-Boltzmann constant, $\sigma_T$ the Thomson scattering
cross section, $\gamma=5/3$ is the ratio of specific heats, and $\ln\Lambda=20$ is the Coulomb
logarithm.

The five differential equations, Eqs.(\ref{e:continuity}), (\ref{e:mdot}), (\ref{e:energy-i}),
(\ref{e:energy-t}), and (\ref{e:fc}), which contain five variables $P(z)$, $T_i(z)$, $T_e(z)$ , $F_c(z)$, and
$\dot m_z(\equiv \rho (z) v_z)$, can be solved with five boundary conditions.  
At the upper boundary, there is no artificial confinement and hence no pressure, which means
a sonic transition at some height $z=H$. As there is no heat flux from/to the boundary, the upper boundary conditions are,
\begin{equation}
F_c=0\  {\rm and}
 \  v_z^2=V_s^2\equiv P/\rho={\Re\over 2\mu}(T_i+T_e)\  {\rm at}\  z=H.
\end{equation}
At the lower boundary of the interface between the disk and corona, the conductive flux
is exactly radiated away and there is no downward heat flux. The temperature of the gas should
be the effective temperature of the accretion disk.  Detailed investigations   \cite[][] {LiuFK1995} show that the temperature increases from the effective temperature to $10^{6.5}$K in a very thin layer and that the conductive flux can be expressed as a function
of pressure at this temperature. Thus, the lower boundary conditions is reasonably
approximated as  \cite[][] {Meyer2000a}
\begin{equation}
T_i=T_e=10^{6.5}K,\ {\rm and} \  F_c=-2.73\times 10^6 P\ {\rm at}\
z=z_0.
\end{equation}
Therefore,  the coronal parameter $P(z)$, $\rho(z)$, $T_i(z)$, $T_e(z)$, evaporation rate $\dot m_z$  can be determined by integrating the differential equation for given mass of black hole, $m\equiv M/M_\odot$, the mass supply rate to the disk, $\dot m\equiv \dot M/\dot M_{\rm Edd}$, and viscosity parameter $\alpha$.
Numerical calculations of the coronal vertical structure revealed that the corona can be approximately divided into two layers, the very thin transition layer with steep variation in temperature and density along $z$, and the geometrically thick, hot corona with decoupled ion and electron temperature. This justifies the configuration of  accretion flows as dominated by   a cold disk and a hot corona connected by a very thin transition layer.    

\subsection{The corona condensation model}\label{s:condensation}
The evaporation occurs when the corona is a tenuous flow. If the density is  sufficiently high so that the radiative cooling in the transition layer  is more efficient than the conductive heating, a certain amount of gas is over-cooled and condenses onto the disk until energy equilibrium  is reached.   Efficient cooling processes in a high-density corona, such as strong inverse Compton scattering of the disk photons, can also facilitate such condensation.  
In the low-mass black hole X-ray binaries, condensation is weak and only occurs in the innermost region since the corona density  built up by evaporation can not be very large.
  In the high mass black hole X-ray binaries, wind captured by the black hole can form a strong corona and condenses to the disk on the way of accretion to the black holes. 
 Such condensation can also occur in AGN accretion flows, accompanied by strong X-ray emission from the efficient comptonization process.
 
For convenient to model the radiation spectrum,  the vertical structure of the disk and corona is simplified to three layers, i.e., the thin disk,  the transition layer, and the corona. 
The parameters in the corona, such as $T_i$, $T_e$, $\rho$, can be determined similar to that of an ADAF, except that  heat conduction and Compton scattering of disk photons are added to the energy balance equation for electrons. Thus, the disk radiation is involved in the corona equations, which depends on the accretion rate in the disk (and coronal radiation if illumination is included).
 Thus,  given the gas supply rate in the outer boundary,  the mass condensation rate can be derived  as the coronal temperatures and density from the energy balance in the transition layer.  This re-allocates the mass accretion rates in the disk and the corona, and in turn, affects the coronal temperatures and density. 
The complete set of equations is list as below. 

The density and pressure in the  corona are,
\begin{equation} \label{e:rho}
\begin{array}{l}
\rho=3.79\times10^{-5}\alpha^{-1}c_1^{-1}c_3^{-1/2}m^{-1}\dot m_c r^{-3/2}~{\rm g~cm^{-3}},\\
n_{\rm e} =2.00\times10^{19}\alpha^{-1}c_1^{-1}c_3^{-1/2}m^{-1}\dot m_c r^{-3/2}{\  \rm cm^{-3}},\\
p =1.71\times10^{16}\alpha^{-1}c_1^{-1}c_3^{1/2}m^{-1}\dot m_c r^{-5/2}{\ \rm g\,cm^{-1}\,s^{-2} },
\end{array}
\end{equation}
where $\dot m_c$ denotes the Eddington-scaled accretion rate in the corona, $m$ is the black hole mass in units of solar mass, $r$ is the radius in units of Schwarzschild radius. $c_1\approx 0.6$ and $c_3\approx0.4$ are insensitive to  the chosen parameters (precisely, $c_1=\frac{5+2\epsilon'}{3\alpha^2}g(\alpha,\epsilon')$,
$\epsilon'={1\over f}\frac{5/3-\gamma}{\gamma-1}$, $\gamma=\frac{8-3\beta}{6-3\beta}$, $g(\alpha,\epsilon')=\left[1+\frac{18\alpha^2}{(5+2\epsilon')^2}\right]^{1/2}-1$,  and $c_3=(2/3)c_1$. $f$ is the fraction of viscously dissipated energy which is advected, $\beta$ is the ratio of gas pressure and total pressure).

The temperatures are determined by the self-similar solution of the sound speed (${p\over \rho}=c_s^2=c_3\Omega_K^2R^2$, see \cite{Narayan1995ApJ...452..710N})
\begin{equation} \label{e:TiTe}
    T_{\rm i}+1.08T_{\rm e}=6.66\times10^{12}\beta c_3r^{-1}{\rm K}
    \label{eq:tite}
\end{equation}
and the energy balance in electrons
\begin{equation} \label{e:energy-e}
    q_{\rm ie}(\rho, T_i, T_e)=q_{\rm rad}(\rho,T_e)+{\Delta F_c/H},
    \label{eq:te}
\end{equation}
where $q_{ie}$ is  the heating rate to electrons (see Eq.\ref{e:qie}),  $\Delta F_{\rm c} /H\approx  k_{0}T_{\rm e}^{7/2}/H^2$ is  an approximation of the conductive cooling rate, 
$q_{\rm rad}$  is the radiative cooling rate  including the  bremsstrahlung, synchrotron,
self-Compton scattering and  external Compton scattering  (see  Eq.\ref{e:comp}), which  are the functions of $T_e$, $\rho$, and $H$  (\cite[e.g.][]{Narayan1995ApJ...452..710N,Manmoto1997ApJ...489..791M})  and  the disk emission. 

The disk is heated up by the viscous stress in the accretion and illumination from the corona, producing radiation with an effective temperature    \cite[][]{liu2011}, $T_{\rm eff}(R)$,  
\begin{equation}\label{e:Teff}
\sigma T_{\rm eff}^4(R)={3GM\dot M_{\rm d}(R)\over 8\pi R^3}\[1-\({3R_S\over R}\)^{1/2}\]
+ {(1-a)L_{\rm c,int} \over 8\pi }{H_s\over (R^2+H_s^2)^{3/2}}
\end{equation}
where $\dot M_{\rm d}(R)$ denotes the accretion rate in the disk,  $L_{\rm c,int}=2\int q_{\rm rad} 2\pi RHdR$ is volume-integrated radiation luminosity from the corona on both sides of the disk,  $H_s\sim (3-10)R_s$ is the height of the coronal illumination  approximated as a lamppost, and  $a$ is albedo of the disk.   Therefore, the soft photon involved in the inverse Compton scattering in the corona is dominated by the local disk radiation in the inner region with an energy density $u=(2/c)\sigma T_{\rm eff}^4$; While in the outer region,  photons  emitted from the inner disk could overwhelm the  local disk emission. Thus,  the energy density $u(R)$ for Compton scattering can be approximated by
\begin{equation}\label{e:softdensity}
u(R)\approx {2\over c}\max \[\sigma T_{\rm eff}^4(R), (1-1/\sqrt{2}){L_d\over 8\pi R^2}\],
\end{equation}
where the second term represents vertically averaged irradiation of disk emission to the outer corona ($L_d$ the total disk luminosity)    \cite[][]{liu2003}.  

From Eqs.(\ref{e:rho}), (\ref{e:TiTe}),(\ref{e:energy-e}), (\ref{e:softdensity}) the temperatures ($T_i$, $T_e$)  and  the density ($\rho$ or $n_e$)  in the corona, and the effective temperature in the disk  can be calculated numerically for given $\alpha$, $\beta$, $a$,  $m$, $R$, provided that the accretion rate in the corona ($\dot m_c$), and in the disk ($\dot M_d$) are known. 

Given the corona temperature and density, the  evaporation rate at unit surface area  can be derived from the transition layer, that is \cite[e.g.][]{liu2007,Meyer2007},
\begin{equation}
    \dot{m_{\rm z}}= \frac{\gamma-1}{\gamma}\beta
    \frac{{-F_{\rm c}}}{ kT_{\rm i}/ \mu_i m_p}(1-\sqrt{C})
	\label{eq:mdotz}
\end{equation}
with
\begin{equation}
    C=\kappa_0 b \left({\frac{\beta^2 p^2}{\pi k^2}}\right)\left(\frac{T_{\rm cpl}}{{F_{\rm c}}}\right)^2,
    \label{eq:ccc}
\end{equation}
where $b=10^{-26.56}{\rm g}\, {\rm cm^{5}}{\rm s^{-3}}{\rm K^{-1/2}}$, $F_{\rm c}\approx  -k_{0}T_{\rm e}^{7/2}/H$, and $T_{\rm cpl}$ the coupling temperature in the transition layer determined by the energy balance  ($q^++q^c=q_{\rm ie}$ with compressive heating rate $q^c={f\over 1-\beta}q^+$) in this layer. $T_i$ and $p$ are the ion temperature and pressure of the corona. 

The positive value of   $\dot{m_{\rm z}}$ means that gas evaporates from the disk to the corona, while  negative $\dot{m_{\rm z}}$ means that gas condenses from the corona to the disk. As a consequence, the accretion rates in the disk and in the corona  vary with the distance. The integrated evaporation/condensation rate (in unit of Eddington rate $\dot M_{\rm Edd}$) from   $R$ to the outer boundary $R_b$ is
\begin{equation}\label{condensation}
\dot m_{\rm evap}(R)= \int_{R}^{R_b} {4\pi R \over \dot M_{\rm
Edd}} \dot m_z dR.
\end{equation}
 If the gas captured by the black hole  is supplied to  disk  at a rate  of $\dot m$, as  the Roche-lobe overflow in the LMXBs,  disk gas will evaporate to the corona when it accretes toward the black hole. The accretion rates  at distance $R$ via the disk and the corona are, respectively,  
 \begin{equation} \label{e:ratedc1}
\dot m_{\rm d}(R)=\dot m -\dot m_{\rm evap}(R)
{\   \  \rm and \  \ } \dot m_{\rm c}(R)=\dot m_{\rm evap}(R).
\end{equation}  
 If the gas is supplied to the corona at a rate of  $\dot m$,  gas can  condense to the disk when the gas supply rate is sufficiently high. We then have
  \begin{equation} \label{e:ratedc2}
  \dot m_{\rm d}(R)= -\dot m_{\rm evap}(R)
{\   \  \rm and \  \ } \dot m_{\rm c}(R)=\dot m+ \dot m_{\rm evap}(R).
\end{equation} 
As the $\dot m_{\rm evap}(R)$ is  an implicit  function of  $T_i$, $T_e$, $\rho$, and $T_{\rm eff}$ for given 
$\alpha$, $\beta$, $a$,  $m$, $R$, $\dot m$,  the accretion rates  in disk and in corona  are  the implicit function of $T_i$, $T_e$, $\rho$, and $T_{\rm eff}$.  Therefore, Eqs.(\ref{e:rho}), (\ref{e:TiTe}),(\ref{e:energy-e}), (\ref{e:Teff}) are a complete set of equations when the  accretion rates in disk and corona are expressed by Eq.(\ref{e:ratedc1}) or Eq.(\ref{e:ratedc2}).   Solving the equations we determine the the disk and corona characteristic parameters $T_i$, $T_e$, $\rho$, and $T_{\rm eff}$. Other quantities, such as the pressure, viscosity heating rate and its advection fraction, the  evaporation/condensation rate, the radiation cooling functions by different processes, are also functions of $T_i$, $T_e$, $\rho$,  $T_{\rm eff}$  and involved in calculating the characteristic parameters.  

\subsection{Consequence of disk-corona interaction: the two-phase accretion flows}\label{s:geometry}
Interaction of the disk and corona results in gas exchange between the two accretion flows, with an evaporation/condensation rate in dependence on the property of the two flows.  The final steady geometry of the accretion flows depends not only on the mass-supply rate, but also on how the gas is supplied to the accretion,  the RLOF or the wind.

\begin{figure}
\begin{center}
\includegraphics[scale=0.3]{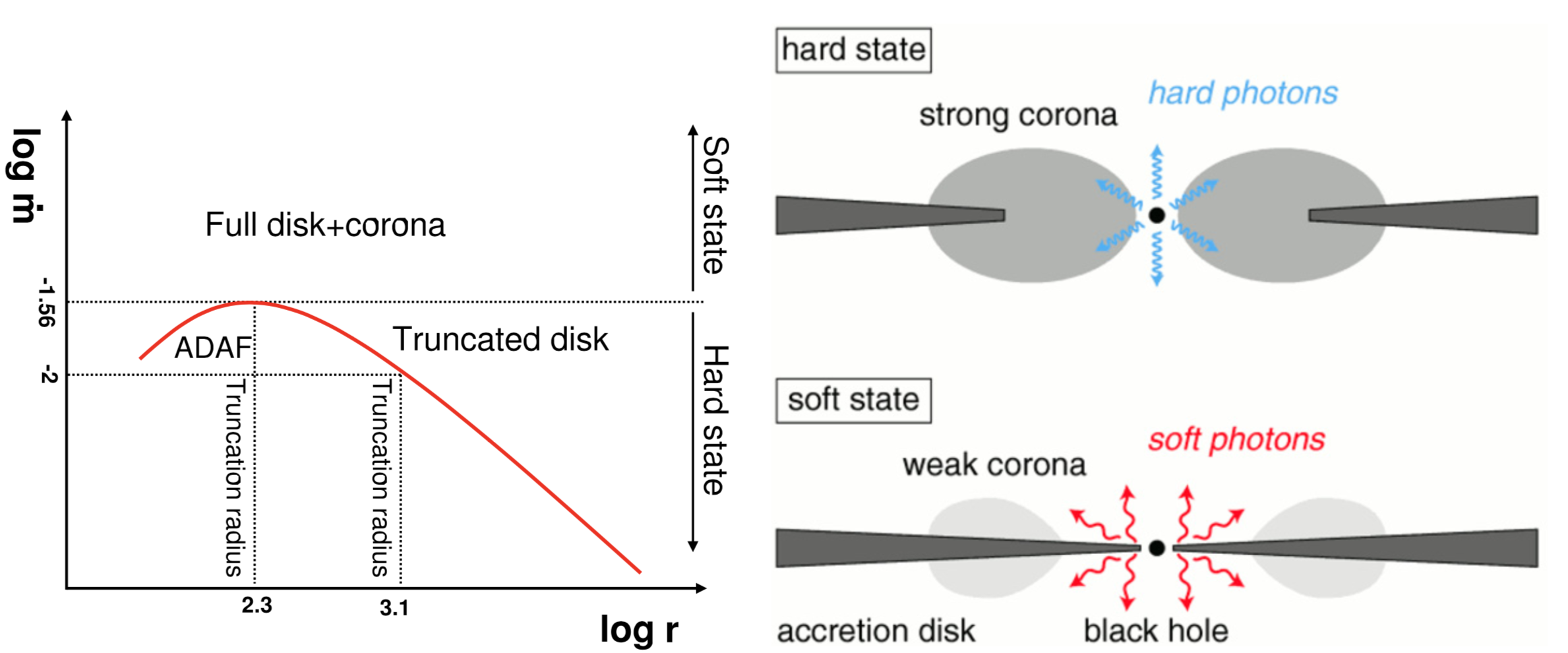}
\caption{Schematic description of disk truncation and the spectral transition illustrating its dependence on the mass accretion rate. The evaporation rate (in units of Eddington rate) is illustrated as a function of the distance (in units of Schwarzschild radius) in the left panel. At low accretion rates supplied by RLOF, the disk is truncated by evaporation due to insufficient mass supply at distances determined by the evaporation curve,  When the RLOF  rate exceeds the maximum, evaporation cannot evacuate any disk region, thus, the disk extends to the ISCO and dominates the accretion flow.  Figure from \cite{liu2009} }
\label{f:disk-corona-evaporation}
\end{center}
\end{figure}

In the case of RLOF supply to the thin disk,  the coronal flow increases during its accretion  toward the black hole as the evaporation gas continuously joins; Meanwhile the radiation efficiency  increases, which depresses  the evaporation from some distance inwards. Thus, the evaporation rate is expected to reach a maximum value and decreases in the innermost region, as shown by numerical computation.
While the evaporation process diverts the accretion flow from the disk to the corona, the disk is completely evaporated  within a specific region for lower gas supply rates.  This region is filled with coronal gas, and the accretion takes place in a geometrically thick, hot flow. On the other hand, for gas supply rate higher than the maximal evaporation rate, the optically thick disk cannot be significantly depleted at any distance. Thus, the disk extends to the ISCO with an overlying accreting coronal flow coexisting with the disk, where continuous evaporation feeds the corona accretion.  The typical accretion geometry is displayed in  Figure \ref{f:disk-corona-evaporation}.

\begin{figure}
\begin{center}
\includegraphics[scale=0.19]{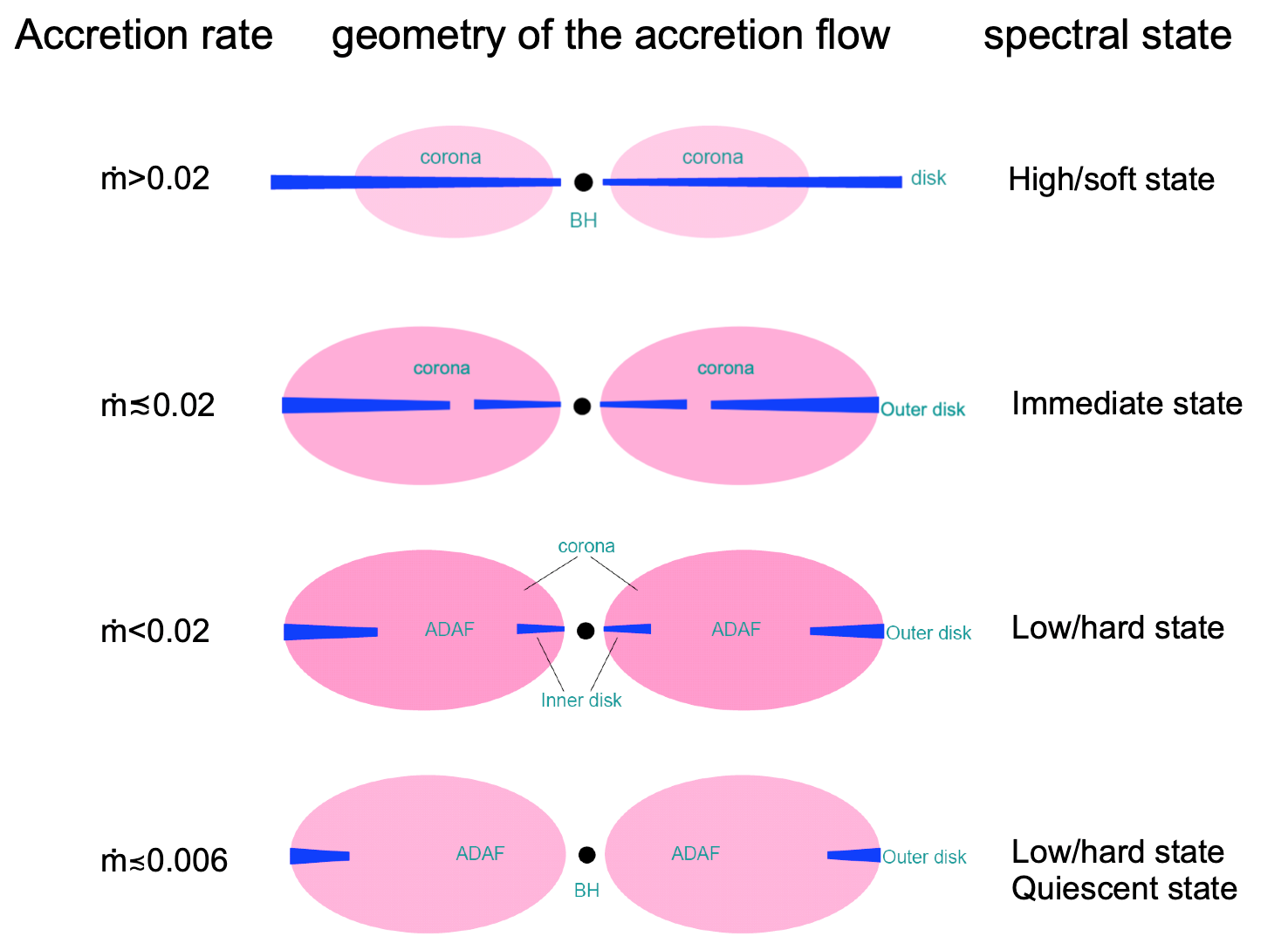}
\caption{Variation of the accretion flow geometry and spectral states with the accretion rate of gas    supplied  to the thin disk, as  the Roche-lobe overflow  in low mass X-ray binaries.  The disk truncation radius moves inwards with increase of gas supply rate as a sequence of equilibrium of evaporation and inflow of gas in the disk; When the gas supply rate is larger than the maximal evaporation rate, the disk cannot be truncated anymore, in stead, the evaporation-fed corona becomes weak as the increased  Compton cooling to the corona restrains the evaporation. An inner disk  occurs during the decay of an outburst since the disk is first depleted at the most efficiently evaporating region around 200 Schwarzschild radii. The inner disk  becomes small with decrease of gas supply  rate and eventually disappears. } 
\label{f:transition-evaporation}
\end{center}
\end{figure}

The evolution of accretion geometry during an outburst of LMXBs is illustrated in Figure \ref{f:transition-evaporation}.  On the  rise of an outburst the accretion rate increases, pushing the truncated disk inwards.  When the critical accretion rate (of $\sim 0.02$) is reached,  the thin disk extends to the ISCO. Further increase of  accretion rate in the disk leads to  efficient Compton cooling and hence condensation of coronal gas. Then the corona becomes very tenuous. During the decay,  the disk weakens with decrease of accretion rate, is truncated at the most efficient evaporation region when  the accretion rate decreases to a lower critical value, leaving an inner disk at the intermediate state; The inner disk eventually disappears  with further decrease of accretion rate.

\begin{figure}
\begin{center}
\includegraphics[scale=0.17]{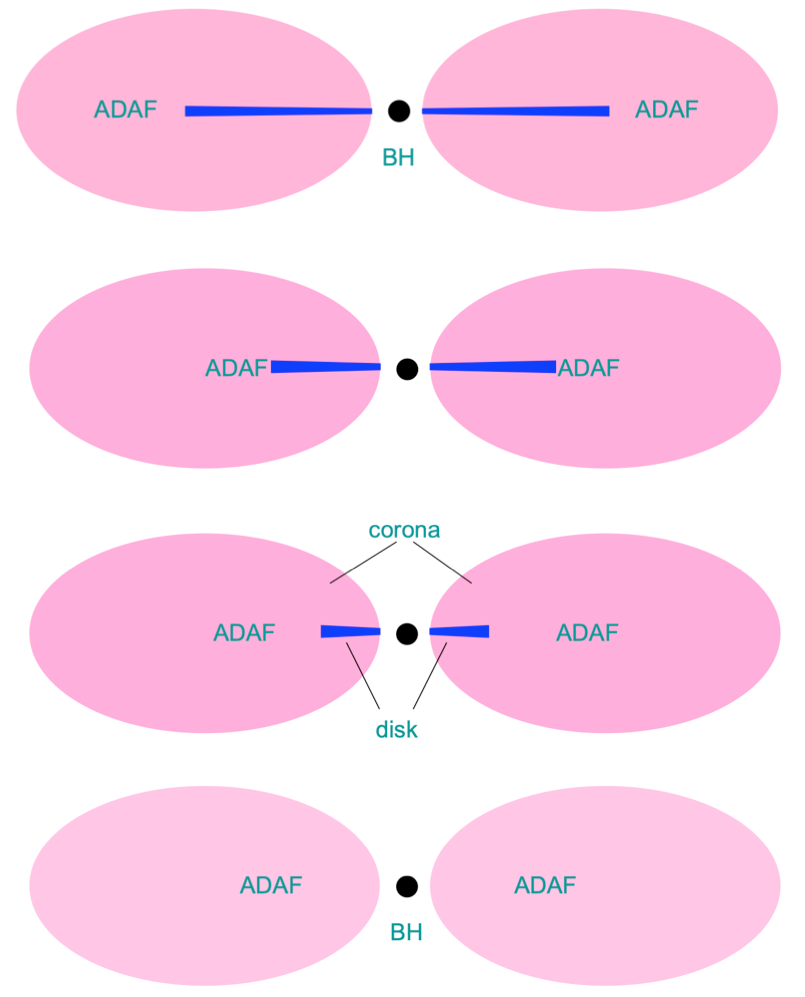}
\caption{Variation of  the accretion flow geometry   with accretion rate of gas   supplied by wind, as that in high mass X-ray binaries or AGNs. The weak wind captured by the black hole forms an ADAF;  When the wind supply rate increases to  a critical value,  part of the corona gas can collapse to form a weak disk in the inner region as a result of over-cooling; Further increase of wind supply leads to strong condensation and a large disk. } 
\label{f:transition-condensation}
\end{center}
\end{figure}

In the case of wind supply,  the spatial extent and thermal state of the accretion flow in the outer regions can significantly differ from the RLOF. As revealed by the 3D hydrodynamical simulations  \cite[][] {Walder2014ASPC..488..141W}, the mass captured from a stellar wind  is heated at the bow shock to temperatures of $\sim10^7$K,  supplied to the accretion in a hot rather than in a cool physical state. Such gas is not necessarily constrained to lie in the disk plane, and hence can form a hot accretion flow as a consequence of inefficient radiative cooling in a geometrically thick flow, i.e., ADAF. By studying such an ADAF interacting with a pre-existing disk,  it is found that the  accretion geometry depends on the accretion rate supplied by the wind or interstellar medium. At low accretion rates,  the gas evaporates from the thin disk, quickly evacuating the thin disk, if existing initially. Thus, the accretion is via  ADAF when it reaches a steady state. This  result  is similar to  that  of cold gas supply to a thin disk, e.g. the RLOF supply.   
At high accretion rate, it is found that
the hot gas partially condenses to the underlying cool disk as it flows toward the black hole.  Such a thin accretion disk can be maintained in a steady state with gas continuously condensing to the disk.  Since the condensation efficiency depends on the gas supply rate, 
 the radial distribution of mass flow rates in the hot and cool components varies with the accretion rate supplied by winds,  in particular,  the thin disk can be maintained exist  only in a limited radial scale in the neighborhood of the black hole. Therefore, the disk size and its strength relative to the corona depends on the wind supply rate. The variation of accretion flow geometry with  wind supply rate is displayed in Figure \ref{f:transition-condensation}.

It is displayed above that the  accretion flow  is a combination of hot and cold  flows, dominated by the ADAF at  low accretion rates, while by the thin disk at high accretion rates. Such a conclusion is valid no matter the accretion gas is supplied by RLOF or by wind/interstellar medium.  This is essential as the hot  gas inevitably interacts with the cold gas during the accretion.  The variation of the gas supply rate  changes the  relative strength of the thin disk and the hot accretion flow,  leading to a change in the spectrum  and  even  state transition, as discussed in the following section.

 
\subsection{The radiative properties as compared with observations}\label{s:radiation}
\begin{figure}
\begin{center}
\includegraphics[scale=0.28]{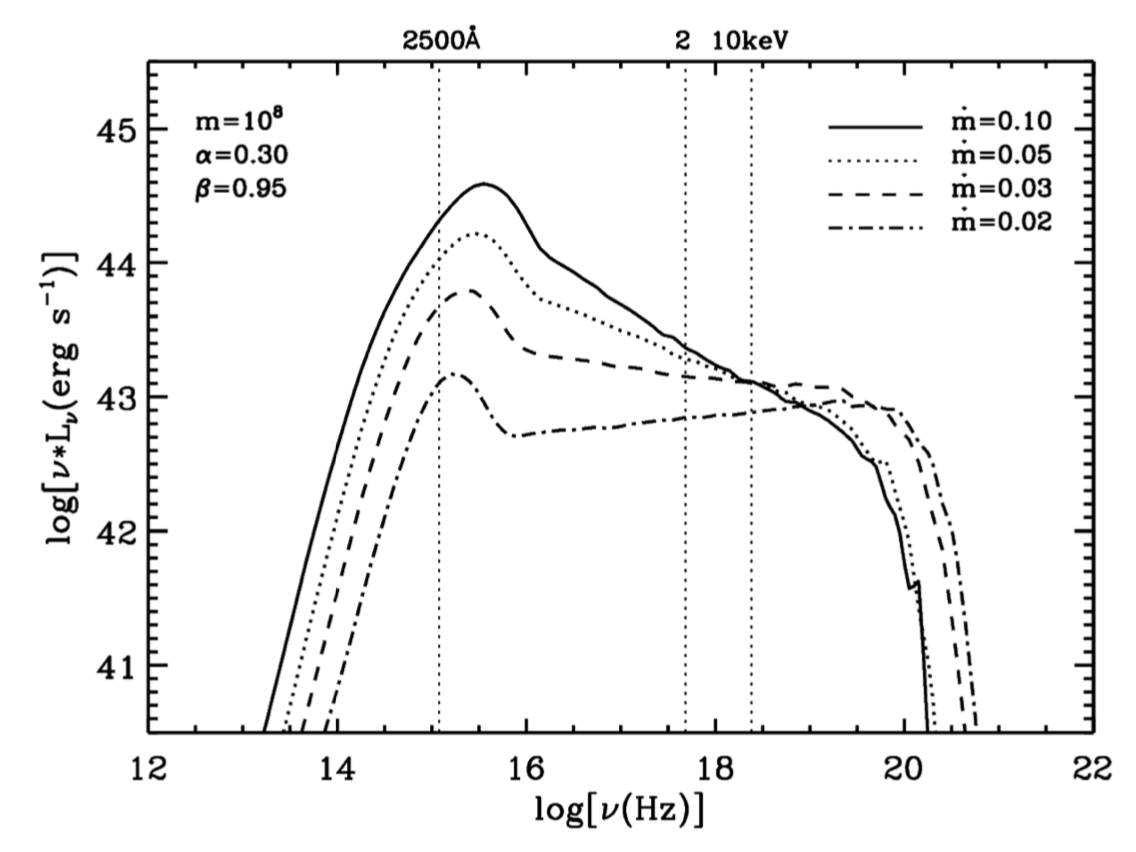}
\includegraphics[scale=0.27]{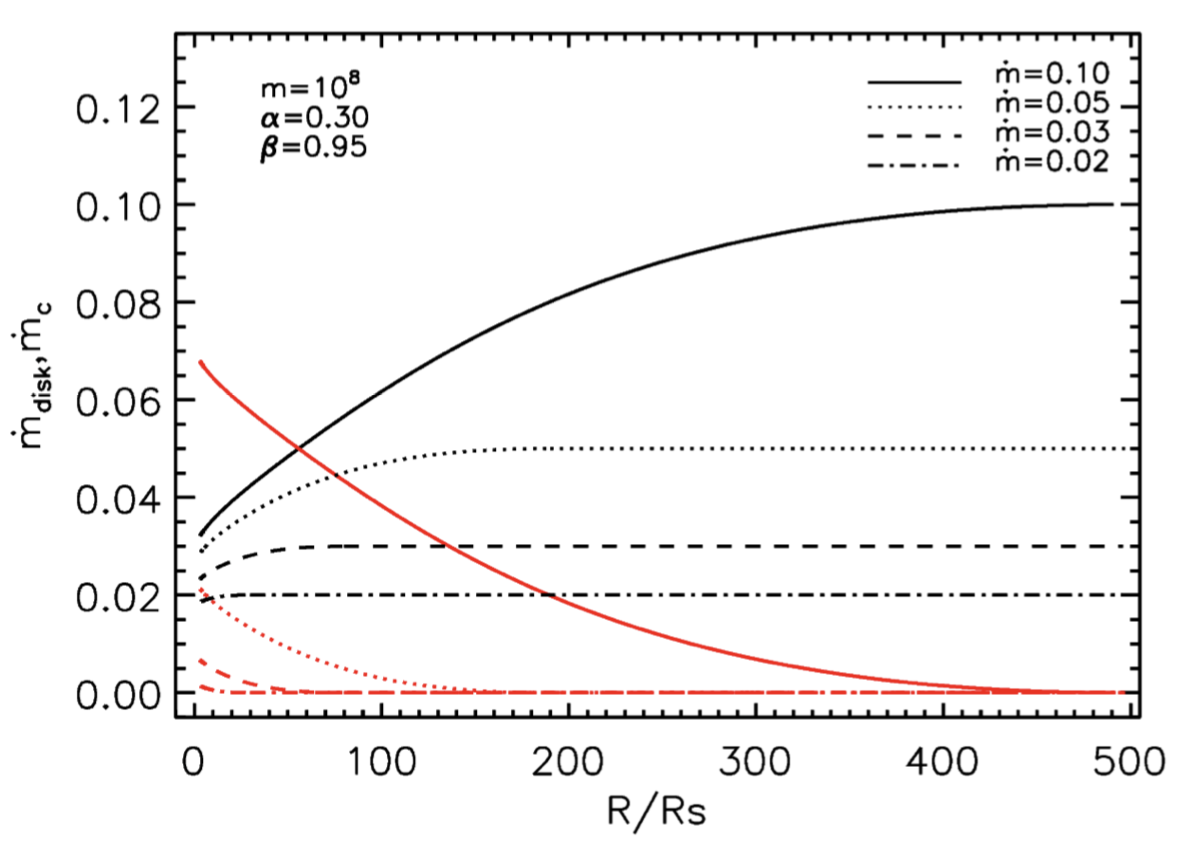}
\caption{Mass flow rate in the disk and corona (right panel) and the corresponding spectra (left panel). Left panel: Spectra emitted from a disk and corona with a hot mass supply for a black hole of $10^8M_\odot$, viscosity parameter $\alpha= 0.3$, magnetic field parameter $\beta = 0.95$, albedo $a=0.15$.  Curves from bottom to top refer to mass supply rates of 0.02, 0.03, 0.05, and 0.1. Right panel: The corresponding radial distribution of the mass flow rate in the disk and corona caused by condensation. Curves in red (black)  refer to the mass flow rate in the disk (corona).  Figure adapted from  \cite[][] {qiao2017} }
\label{f:spectra}
\end{center}
\end{figure}

The emission of accretion flows comes dominantly  from the inner region, as a results of release of gravitational energy into radiation.  Given the fact that the radiation efficiency  of hot gas increases toward the black hole (as a consequence of efficient Coulomb coupling at inner region), the corona (or ADAF) emission is even more dominated by the inner region  \cite[][] {liu2017},  no matter  the corona  is compact or extended.  Therefore, the total radiation and SED from the disk and corona are not significantly affected by the style of gas supply, but are determined by  accretion rates flowing respectively in  the inner disk and inner corona, which are eventually determined by the  the gas supply rate.  An example of spectral variation with the gas supply rate from the accretion flows is displayed in Figure \ref{f:spectra} for the case of wind accretion. 
The left panel of the figure shows that  the spectra are soft at high accretion rate, with strong  disk emission and steep X-ray spectrum.  This is caused by the strong condensation of coronal gas to the disk. When the  accretion rate decreases,  condensation becomes weak,  thereby the accretion rate in the disk decreases, as shown in the right panel of  Figure \ref{f:spectra}.  Condensation stops when the gas supply rate decreases to $\sim 0.02$.   Correspondingly, the spectrum undergoes transition from soft state to hard state.  

\begin{figure}
\begin{center}
\includegraphics[scale=0.22]{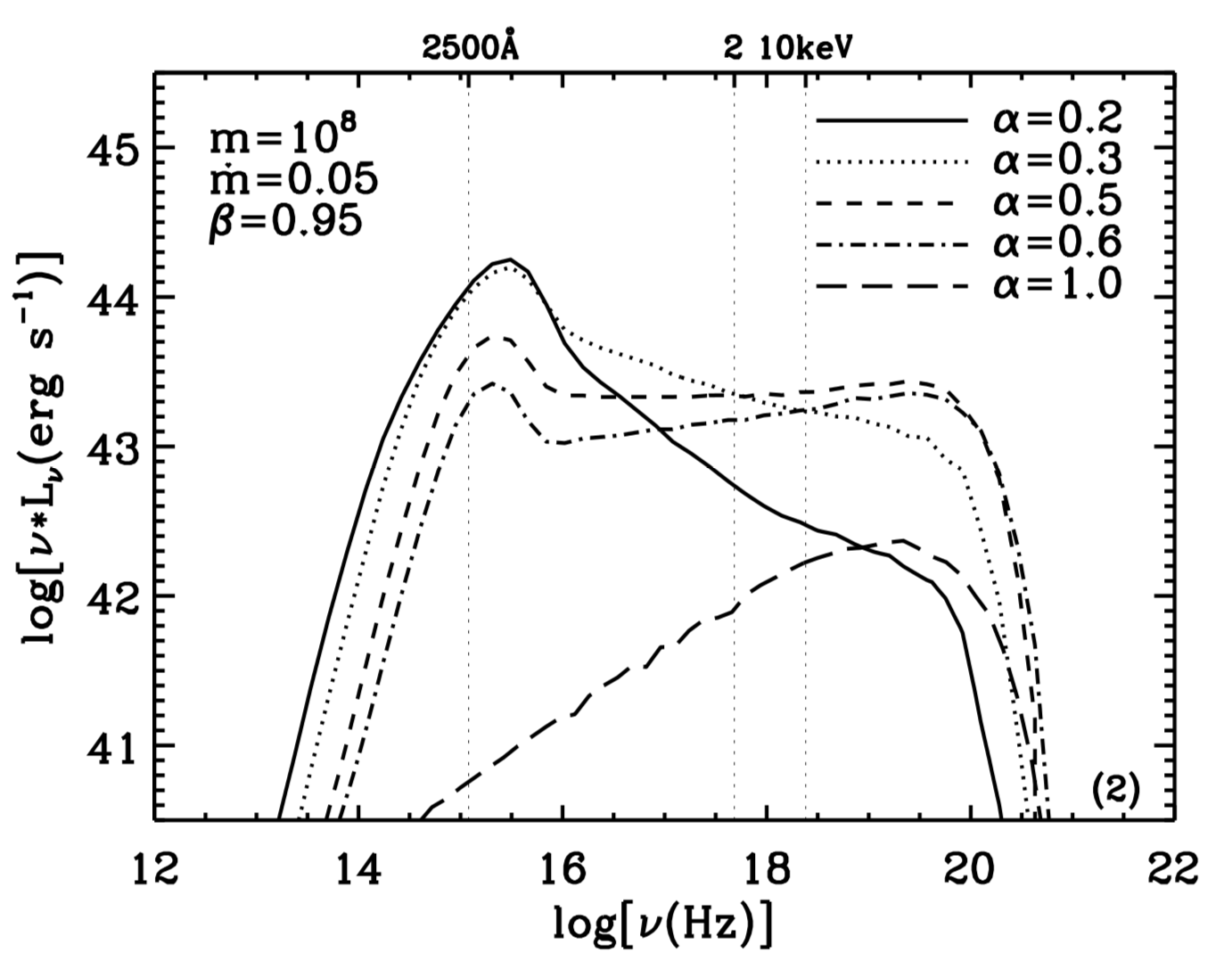}
\includegraphics[scale=0.22]{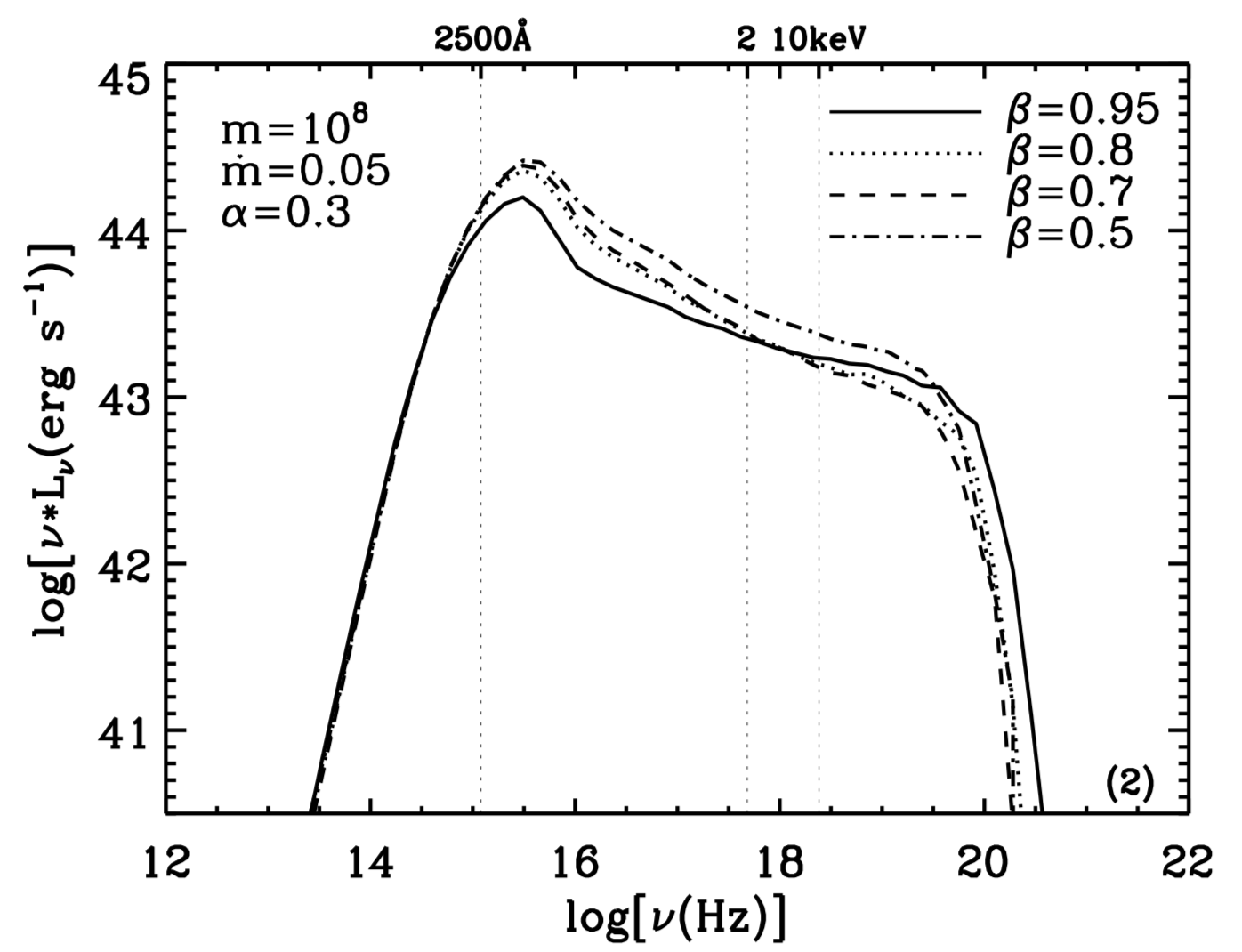}
\caption{Spectra dependent on the viscosity (Left panel) and magnetic field (Right Panel).  Figures from  \cite[][] {qiao2018a} }
\label{f:spectra-parameter}
\end{center}
\end{figure}

We note that the  relative strength of disk and corona, and hence the emission spectrum,  also depend on the viscosity parameter ($\alpha$), the magnetic field (parameterized in $\beta$ ), and the albedo ($a$) of the disk  \cite[][] {qiao2018a}. A larger value of $\alpha$ means more heating, leading to a stronger corona and harder X-ray spectrum; The spectrum is not sensitive to the magnetic field if the magnetic energy is no larger than the energy of equipartition (i.e. $0.5<\beta<1$); The effect of albedo is also not very important since it  varies  only in a limit range, for example, $a=0.1-0.2$ for a disk around AGN  \cite[][] {Haardt1993}.  Nevertheless, a change in the combination of the three parameters can,  to some extent, change the geometry of the two-phase accretion flows, thereby change the overall spectral shape.  The effects of viscosity and magnetic field on the spectra  for $a=0.15$ are shown in Figure \ref{f:spectra-parameter}.

Therefore, the two-phase accretion flows can produce various types of spectra, which are mainly determined by the gas supply rates, modified by the value of viscosity as well as magnetic filed and albedo.  The transition of spectral states can occur at an accretion rate depending on the viscous parameter and evolution history (from hard to soft state or from soft to hard state). The model naturally interprets the diversity of spectra and state transition of black holes.     

While the general features of the stationary accretion flows are similar for the accretion supplied by RLOF and wind,  there exist some differences.  In contrast to the extensive thin disk formed from RLOF,   the disk  is much smaller as fed by the condensation of hot wind because the radiative cooling at outer region is insufficient to cause condensation. The absence of an  outer disk precludes the possibility of thermal instabilities in the hydrogen ionization zone and therefore the  large outbursts.  This interprets why  Cyg X-1 is  a persistent source with no large outburst, unlike the transient sources. Luminosity fluctuations in the  wind-fed sources would be a consequence of turbulence in the accretion wake region, which are likely smaller than that caused by thermal instabilities. Thus, the wind-fed sources could be described as persistent, varying occasionally between soft and hard spectral states with a relatively small amplitude.

The second difference is  the presence/absence of hysteresis  between the transition luminosities at hard-to-soft and soft-to-hard transitions. The hysteresis in the light curve of transient sources \cite[e.g.][]{Maccarone2003MNRAS.338..189M, Zdziarski2004PThPS.155...99Z} is thought to be caused by the large  difference in Compton cooling rate  before the state transitions
  \cite[][] {liu2005,Meyer2005a}. When the transient sources are in a deep quiescent state before outbursts are triggered, the evaporation is efficient as there is no/weak Compton cooling in the corona. Only when the accretion rate exceeds the maximal evaporation rate can the disk is fully filled in and spectral state transits.  During the decay of an outburst the strong disk radiation involved in Compton scattering depresses the disk evaporation, which leads to the disk to be truncated  at a lower accretion rate.  However, this effect is insignificant in persistent sources because of a small difference in the variation of accretion rate in the hard and soft states  \cite[][] {Taam2018}. Therefore, the light curve during the transition between states is likely to be more symmetrical in persistent systems, as it is observed in Cyg X-1. 

 The third distinction  is that the accretion rates remained in the innermost corona are similar ($\dot m_{\rm c} \sim 0.02$) for different rates of the hot gas supply ($\dot m >  0.02$). This is in contrast to the conclusion from RLOF that very little gas accretes via a corona  at a high gas supply rates, i.e. the higher gas supply rate,  the lower rate in the corona. Therefore, the X-ray emission in wind accretion systems, which originates from the liberation of gravitational energy of the coronal gas and enthalpy of the condensed gas, can be larger than that with  only cold gas supply to the disk (e.g. RLOF). Hence, the wind accrretion model provides insight to the generation of strong X-ray radiation in luminous AGNs without necessarily invoking artificially  heating to the corona. 

Finally, there should be difference in timing properties between the accretion of RLOF and wind, if spatial region of hot and cold gas is involved in producing the noise component. 

\section{Conclusion}
This paper reviews the physics of formation of multi-phase accretion flows around black holes. Of importance is the fact  that the coupling between the cold and hot gas leads to gas evaporation or condensation, thereby the formation of various configurations of accretion flows as a function of accretion rate and system parameters. In particular,  it shows how the gas supply affects the accretion configuration. For the Roche-lobe overflow supply, as in low mass X-ray binaries,  evaporation from the disk leads to accretion via  a thin disk sandwiched in a corona at high accretion rates, and an inner ADAF connecting to an outer disk at low accretion rates;  For the wind or interstellar medium supply,  an ADAF forms and extends to ISCO at low accretion rates, while a thin disk forms in the inner region owing to hot gas condensation,  which is coexisted with the left corona at high accretion rates.   The different configuration caused by different  gas supply does not significantly affect the radiation spectrum, except for the relatively strong coronal emission in the case of wind supply. However, it does cause different variability features, such  as the presence/absence of hysteresis in transition luminosity,   the transient  behavior in LMXBs and persistent  behavior  in Cyg X-1.

{\bf Acknowledgments:} Financial support for this work is provided by the National Natural Science Foundation of China (grants 12073037 and 11773037) and NAOC Nebula Talents Program.

\bibliography{bibfile}

\end{document}